%% file: paper_183.tex
\documentclass[12pt]{article}
\usepackage{a4,epsfig}

\newcommand{\sqs}{\ensuremath{\sqrt{s}}}
\newcommand{\mvis}{\ensuremath{m_{\mathrm{vis}}}}
\newcommand{\mh}{\ensuremath{m_{\mathrm H^\pm}}}
\newcommand{\mw}{\ensuremath{m_{\mathrm W}}}
\newcommand{\mhr}{\ensuremath{m^{\mathrm{rec}}_{\mathrm H}}}
\newcommand{\mwr}{\ensuremath{m^{\mathrm{rec}}_{\mathrm W}}}
\newcommand{\hpm}{\ensuremath{\mathrm H^{\pm}}}
\newcommand{\hh}{\ensuremath{\mathrm{H^+H^-}}}
\newcommand{\btn}{\ensuremath{{\mathrm{B}(\mathrm{H}^+\!\to\!\tau^+\nu_\tau)}}}
\newcommand{\tntn}{\ensuremath{\tau^+\nu_{\tau}\tau^-\bar{\nu}_{\tau}}}
\newcommand{\tncs}{\ensuremath{\mathrm{c\bar{s}}\tau^-\bar{\nu}_{\tau}}}
\newcommand{\cstn}{\ensuremath{\mathrm{\bar{c}s}\tau^+\nu_{\tau}}}
\newcommand{\cscs}{\ensuremath{\mathrm{c\bar{s}s\bar{c}}}}
\newcommand{\ee}{\ensuremath{\mathrm{e^+e^-}}}
\newcommand{\ww}{\ensuremath{\mathrm{W^+W^-}}}
\newcommand{\qq}{\ensuremath{\mathrm{q\bar{q}}}}

\newcommand{\gevc}{\ensuremath{\mathrm{GeV}/c}}
\newcommand{\gevcc}{\ensuremath{\mathrm{GeV}/c^2}}
\newcommand{\invpb}{\ensuremath{\mathrm{pb^{-1}}}}

\newcommand{\nexb}{\ensuremath{N^{\mathrm{exp}}_{\mathrm{bkg}}}}
\newcommand{\nsub}{\ensuremath{N^{\mathrm{sub}}_{\mathrm{bkg}}}}
\newcommand{\nobs}{\ensuremath{N_{\mathrm{obs}}}}

\font\ninerm=cmr9

\begin{document}

\thispagestyle{empty}

\begin{titlepage}

\begin{picture}(160,1)
\put(2,20){\rm\large EUROPEAN LABORATORY FOR PARTICLE PHYSICS (CERN)}
\end{picture}

\begin{center}
\vspace{2.1cm}
{\Huge Search for charged Higgs bosons\\
in ${\mathrm e}^+{\mathrm e}^-$ collisions 
at \sqs\ = 181--184~GeV }
\vspace{1.8cm}

The ALEPH Collaboration$^*)$ 

\end{center}

\vspace{2.2cm}
\begin{abstract}
\vspace{.5cm}
Data collected at centre-of-mass energies of   
181--184~GeV by {\tt ALEPH} at
{\tt LEP}, \linebreak corresponding to an integrated luminosity of
56.9~\invpb, are analysed in a search for pair-produced
charged Higgs bosons~\hpm.
Three analyses are employed to select the \tntn, \tncs{}/\cstn\ and \cscs{}
final states. 
No evidence for a signal is found. Mass limits are set as a
function of the branching 
fraction~\mbox{\btn}.
Under the assumption that the decay modes considered cover the 
totality of the possible final states, 
charged Higgs bosons with masses below 59~\gevcc{} are 
excluded at $95\%$~C.L. 
independently of ~\mbox{\btn}.
\end{abstract}
\vfill
\vskip .5cm
\noindent
--------------------------------------------\hfil\break
{\ninerm $^*)$ See next pages for the list of authors}
\end{titlepage}

\clearpage
\include{authb}

\pagestyle{plain}
\pagenumbering{arabic}

\section{Introduction}

Despite the success of the Standard Model of electroweak interactions 
in describing experimental observations, not much information is 
available about its cornerstone, the Higgs sector. 
In its minimal version, the Higgs mechanism is implemented 
by adding only one doublet of complex scalar fields, resulting 
in one additional physical scalar state, electrically neutral, 
commonly referred to as the standard Higgs boson.
The most important phenomenological consequence of 
an extended Higgs structure is the appearance of 
additional physical spin-0 states~\cite{HUNTERS}. 
For example, with the addition of one more doublet of complex
scalar fields, five physical states remain
after spontaneous breaking of the $\mathrm{SU(2)_L\times U(1)_Y}$ 
symmetry to give 
mass to $\mathrm{W}^\pm$ and $\mathrm{Z}$ gauge bosons: three
neutral and a pair of charged bosons.
Among the possible choices, multi-doublet models 
are theoretically interesting because 
they automatically lead, at tree level, to  
$m_{\mathrm W}=m_{\mathrm Z}\cos\theta_{\mathrm W}$ 
and to the absence of 
flavour changing neutral currents, two major constraints 
which must be satisfied by any extension of the Standard Model to agree
with the experimental observations.

This letter describes a search for pair production 
in $\ee$ collisions 
of the charged Higgs bosons \hpm{} predicted in 
two-Higgs-doublet extensions of the Standard Model. 
The analysis uses the total integrated luminosity of 56.9~pb$^{-1}$ 
collected in 1997 with the {\tt ALEPH} detector at {\tt LEP},  
at centre-of-mass energies from 181 to 184 GeV, hereafter 
called the 183 GeV data. 

Pair production of charged Higgs bosons occurs mainly via 
$s$-channel exchange of a photon or a Z boson; in two-doublet models, 
the couplings are completely specified in terms of the electric 
charge and $\theta_{\mathrm W}$, making
the production cross section depend only on one additional 
parameter, the charged Higgs boson mass~\mh.
As expected in most implementations of multi-doublet 
models~\cite{HUNTERS}, 
it is assumed that $\mathrm{H^+}$ decays, with negligible lifetime,
predominantly into $\mathrm{c\bar{s}}$ or ${\mathrm \tau^+\nu_\tau}$ 
(and the respective charge conjugates for ${\mathrm H}^-$). 
Additional decay channels, such as those involving neutral Higgs bosons, 
are not considered here.
Since the relative weight of the two main channels depends on the 
details of the model, no assumption is made about the decay
branching fractions, 
and three different selections are developed to address the 
possible final states \cscs{}, 
\tncs{}/\cstn\ (hereafter referred to as \tncs{}) and \tntn. 
Under the same hypothesis, the negative results of the searches performed 
using 27.5~pb$^{-1}$ collected at centre-of-mass energies 
ranging from 130 to 172 GeV 
allowed {\tt ALEPH} to exclude charged Higgs boson masses
less than 52~\gevcc{} at 95\%~C.L., independently of the final
state~\cite{HCH172}.
Using the data recorded at the same centre-of-mass energies, 
an excluded domain up to 54.5~\gevcc\ has also been reported by 
{\tt DELPHI}~\cite{DELPHI172}. 
Charged Higgs boson masses up to  
57.5~\gevcc\ and 59.5~\gevcc\ have been excluded by {\tt L3}~\cite{L3183} and 
{\tt OPAL}~\cite{OPAL183}, respectively, using their data recorded at 
centre-of-mass energies up to 183~GeV.
Less general limits have also been set by
{\tt ALEPH}~\cite{btau}, {\tt CLEO}~\cite{cleo} and {\tt CDF}~\cite{cdf.higgs}.

The letter is organized as follows. 
After the description of the relevant parts of the {\tt ALEPH} detector in 
Section~2, 
the event selections are detailed in Section~3. The results and 
the conclusions are given in Sections~4~and~5.

\section{The ALEPH Detector}
The {\tt ALEPH} detector is described in detail in
Ref.~\cite{bib:detectorpaper}. An account of the performance of the
detector and a description of the standard analysis algorithms can be
found in Ref.~\cite{bib:performancepaper}. Here, only a brief
description of the detector components and of the algorithms relevant for
this analysis is given.

In {\tt ALEPH}, the trajectories of charged particles are measured with a silicon
vertex detector, a cylindrical drift chamber, and a large time
projection chamber. These are immersed in a 1.5~T axial field provided
by a superconducting solenoidal coil.
The electromagnetic calorimeter, placed between the tracking system and the coil, 
is a highly segmented sampling calorimeter which is used to identify electrons
and photons and to measure their energies.
The luminosity monitors extend the calorimetric coverage 
down to 34~mrad from the beam axis. 
The hadron calorimeter consists of the iron return yoke of the magnet 
instrumented with streamer tubes. It provides a measurement of hadronic energy 
and, together with the external muon chambers, muon identification.

The calorimetry and tracking information are combined 
in an energy flow algorithm, classifying a set of 
energy flow ``particles'' as photons, neutral
hadrons and charged particles. Hereafter, charged particle tracks
reconstructed with at least four hits in the TPC,
and originating from within a cylinder of length~20~cm and radius~2~cm
coaxial with the beam and centred at the nominal collision point, 
are referred to as {\it good tracks}.

\section{Event selections}

In order to ensure a good discovery potential independent of the branching
fraction \linebreak \btn{}, three
selection procedures are designed for the
topologies \tntn, \tncs{} and \cscs.
As in Ref.~\cite{HCH172}, the most relevant selection 
criteria for the three selections
are chosen in order to achieve, on average and in case
no signal is present, the best $95\%$ C.L.
limit on the \hh\ production cross section.
To do so, each selection is optimized individually 
with the most optimistic~\btn{} in each case~($100\%$, $50\%$ and
$0\%$ for the \tntn{}, \tncs{} and \cscs{} channels, respectively,
for which the combined contribution of the other two analyses is minimal), 
following the prescription of Ref.~\cite{nbar95} modified to 
include the possibility of partial or full background subtraction. 
In the following sections, the Monte Carlo samples used in designing the 
selections are described and the changes with respect to the 
analyses published in Ref.~\cite{HCH172} are presented. 
In each case, the {\it subtractible} background, i.e., that for 
which the theoretical knowledge and the simulation accuracy are considered 
to be under control, is estimated together with the related systematic 
uncertainty.

\subsection{Monte Carlo Samples}

Fully simulated Monte Carlo event samples reconstructed with the same program
as the data have been used for background estimates, design of
selections and cut optimization.
Samples of all background sources corresponding to at least 
20 times the collected luminosity were generated.
The most important background sources are $\ee\to\tau^+\tau^-$, \qq{}
and four-fermion processes (including \ww production), simulated with
{\tt KORALZ}~\cite{WAS1}, {\tt PYTHIA}~\cite{PYTHIA} and 
{\tt KORALW}~\cite{KORALW}.

The signal Monte Carlo events were generated using the 
{\tt HZHA}~\cite{JANOT} generator, 
extended for charged Higgs boson production as described in Ref.~\cite{HCH172}.
Samples of at least 1000 signal events were simulated for each of the
various final states for 
charged Higgs boson masses between 40 and 80~\gevcc.

\begin{boldmath}
\subsection{The \tntn{} final state}
\end{boldmath}
The final state produced by leptonic decays of both charged Higgs
bosons consists of two acoplanar $\tau$'s and missing energy carried away
by the neutrinos. Since this topology is very similar to that of 
stau pair production,
the selection described in Ref.~\cite{slepton183} 
is used here to search for charged Higgs bosons in the
\tntn{} channel. 
This selection exploits the fact that the signal events contain at least
four neutrinos, leading to large missing energy and a large acoplanarity
of the visible system. Background from \ww production followed by
leptonic W decays is suppressed
by vetoing events with energetic electrons or muons, which are 
softer when originating from $\tau$ decays.

Efficiencies to select events from $\hh\to\tntn$ are of the order of $45\%$, 
as shown in Table~\ref{effbgtable}
for a representative set of Higgs boson masses. The total background expected
amounts to 6.5 events, consisting mainly of 
irreducible background from $\ww\to\tntn$.
In the data, four events were selected, in good agreement with 
the Standard Model expectation. For the interpretation of the negative result of the
search in terms of mass limits, the part of the expected background coming 
from \ww\ is subtracted; the latter amounts to 5.0 events, including 
a reduction of 3\% to account for systematic uncertainties~\cite{slepton183}.

\begin{table}[h]
\begin{center}
\caption{\label{effbgtable}{\small Efficiencies $\epsilon$ (in \%), 
numbers of Standard Model background events expected (\nexb)\  
and subtracted (\nsub), and numbers of observed candidates (\nobs)\ 
 for the three analyses at the
centre-of-mass energy of 183~GeV, as functions of the charged Higgs 
boson masses~(in~\gevcc). 
For the mixed and four--jet channels, numbers are quoted 
within the windows defined by the sliding cuts, therefore implying 
some overlap among different charged Higgs boson mass hypotheses.
}}
\vspace{.3cm}
\renewcommand{\arraystretch}{1.1}
\begin{tabular}{|c||c|c|c|c||c|c|c|c||c|c|c|c|} 
\hline
            & \multicolumn{12}{|c|}{Final state} \\
\cline{2-13}
 \mh      & \multicolumn{4}{|c||}{\tntn{}} & \multicolumn{4}{|c||}{\tncs{}} & \multicolumn{4}{|c|}{\cscs{}} \\
\cline{2-13}
          & $\epsilon$ & \nexb  & \nsub  & \nobs & 
            $\epsilon$ & \nexb  & \nsub  & \nobs & 
            $\epsilon$ & \nexb  & \nsub  & \nobs  \\
\hline
$50$ & 40 & 6.5 & 5.0 & 4 & 37 & 1.2 & 1.0 & 2 & 37 & 5.6  &  5.0 &  5 \\
$55$ & 42 & 6.5 & 5.0 & 4 & 34 & 1.3 & 1.0 & 1 & 36 & 6.8  &  6.1 &  4 \\
$60$ & 43 & 6.5 & 5.0 & 4 & 31 & 1.5 & 1.2 & 2 & 35 & 7.9  &  7.0 & 11 \\
$65$ & 45 & 6.5 & 5.0 & 4 & 25 & 2.1 & 1.7 & 3 & 33 & 8.8  &  7.9 &  9 \\
$70$ & 46 & 6.5 & 5.0 & 4 & 20 & 3.1 & 2.5 & 6 & 32 & 10.5 &  9.4 & 14 \\
$75$ & 48 & 6.5 & 5.0 & 4 & 18 & 5.1 & 4.1 & 5 & 31 & 19.1 & 17.0 & 20 \\
\hline
\end{tabular}
\renewcommand{\arraystretch}{1.}

\end{center}
\end{table}

\begin{boldmath}
\subsection{The \tncs{} final state}
\end{boldmath}

The mixed final state, \tncs{}, 
is characterized by 
two jets originating from the hadronic decay of one of the charged Higgs
bosons 
and a thin $\tau$ jet plus missing energy due to the neutrinos
from the decay of the other. 
\par
Two complementary approaches are used to select the mixed final state:
in one selection, called the {\it global} analysis, global quantities
such as acoplanarity, thrust, and missing momentum are predominantly used
whereas the second selection, referred to as the {\it topological} analysis,
relies more on the specific $\tau$ jet reconstruction.
As the analyses are described in detail in Ref.~\cite{HCH172}, the focus
here is on changes other than a simple rescaling of the cuts 
with \sqs{}. 

As \mh\  approaches \mw\ the sensitivity 
of both analyses is limited by the \ww\  background. 
Two changes are introduced to improve the rejection of this background.
The first change concerns the momentum of the leading
lepton~(electron or muon) which is now required to be less 
than 24~\gevc\  in both analyses.
\par
The second change involves the rejection of the 
$\tau\nu\mathrm{q\bar{q}}'$ final state of W pair events:
cuts are introduced which depend on the signal mass hypothesis ({\it sliding} cuts) 
and are tightened with increasing \mh.

\begin{figure}
\begin{center}
\epsfig{figure=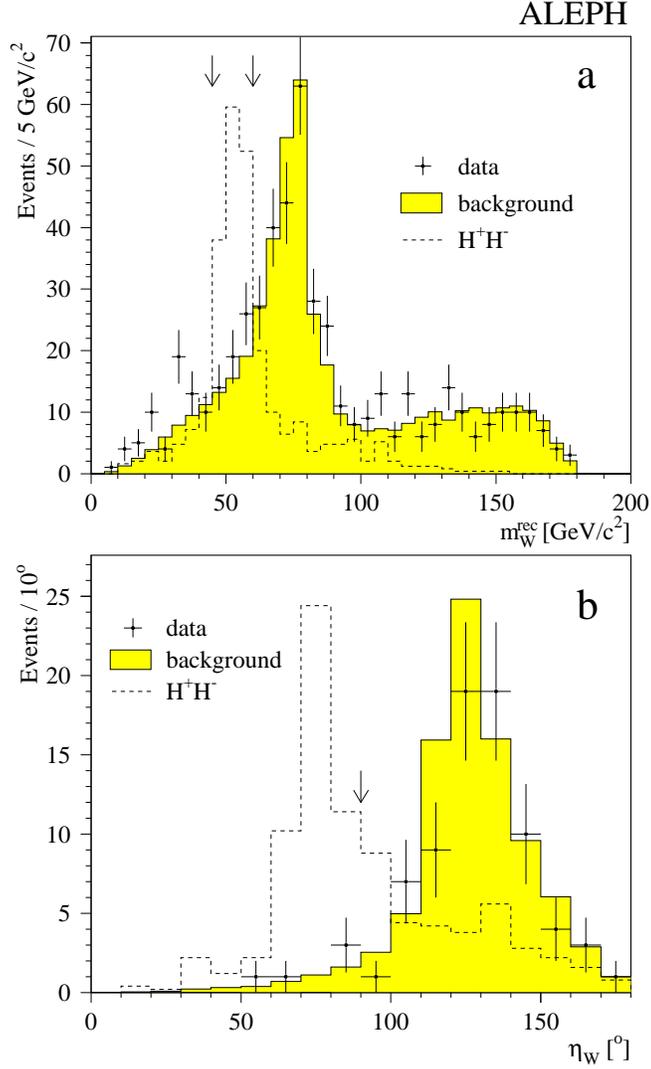,width=0.6\textwidth}
\caption{\label{Fig:hphm_fig1}
{\small Distribution of (a) the invariant mass of the quark jets and 
(b) their acollinearity angle 
used in the \tncs{}\ global selection. 
The dots are the 183~GeV data, 
while the shaded histogram 
is the background expectation, normalized to the recorded
luminosity. The dashed line is the distribution for a 
signal with a Higgs boson mass of 55~\gevcc{}, arbitrarily
normalized. The arrows indicate the position of the cuts applied for this specific  
choice of the charged Higgs boson mass hypothesis (see text for details). 
Only a subset of the cuts is applied here to preserve sufficient 
statistics.}}
\end{center}
\end{figure}

In the global analysis the rejection is achieved with  
cuts on the acollinearity angle $\eta_\mathrm{W}$ of the two hadronic 
jets and their invariant mass \mwr. 
The acollinearity angle is required to be less than 
$[35\!+\!\mh/(\gevcc) ]$~degrees. 
The event is rejected if the
invariant mass lies outside the range of 
$[\mh\!-\!10\ \gevcc ,\mh\!+\!5\ \gevcc ]$.
The good agreement of data and Monte Carlo
in these two variables is shown in Fig.~\ref{Fig:hphm_fig1}.

In the topological analysis the invariant mass \mhr\ 
of the two jets assigned to the Higgs boson must lie
in a window between 
$\mh\!-\!15\ \gevcc$ and $\mh\!+\!5\ \gevcc$.
Their acollinearity $\eta_H$ must be less than
$[40\!+\!\mh/(\gevcc)]$~degrees.

\begin{table}[tp]
\begin{center}
\caption{\label{Tab:tncscuts} {\small 
Summary of the cuts applied in the \tncs{} analyses. 
Units for masses and momenta are \gevcc\ and 
\gevc, respectively. Variables not defined in the text have 
the same meaning as in Ref.~[2]. 
}}
\vspace{.3cm}
\begin{tabular}{|c|c|} 
\hline
          \multicolumn{2}{|c|}{Preselection}  \\
\hline
 Good Tracks & $\geq$7  \\
 Visible Mass~\mvis & [40,$\sqrt{s}$] \\
 Energy below 12$^\circ$ & $< 2.5$\%\sqs \\
 Boost          & $> 0.3$ \\
 $N^{\mathrm{y=0.001}}_{\mathrm{jet}}$ & $\geq 3$ \\
\hline\hline
          \multicolumn{2}{|c|}{Global analysis}  \\
 \hline
 Acoplanarity & $<175^\circ$  \\
 Thrust & $<0.9$  \\
 $E_{\mathrm{wedge}}$ & $<7.5$\%\sqs  \\
 $P_{\mathrm T}$ & $>20\%E_{\mathrm{vis}}$  \\
\hline
 $P_{\mathrm{e},\mu}$ & $<24$ \\
\cline{2-2}
 $m^{\mathrm{NO\ e},\mu}_{\mathrm{vis}}$ & $<80$  \\
\hline
 Tau identification & Loose \\ 
\hline
 $\eta_{\mathrm W}$ & $<[35\!+\!\mh ]^\circ$ \\
 \mwr & $[\mh\!-\!10,\mh\!+\!5]$ \\
\hline\hline
          \multicolumn{2}{|c|}{Topological analysis}  \\
\hline
 $\theta_{\mathrm{miss}}$ & [25.8$^\circ$,154.2$^\circ$]  \\
 $E_{\mathrm{wedge}}$ & $<20$\%\sqs  \\
\hline
 $P_{\mathrm{e},\mu}$ & $<24$   \\
\cline{2-2}
 $m^{\mathrm{NO\ e},\mu}_{\mathrm{vis}}$ & $<80$  \\
\hline
 Tau identification & Tight \\ 
\hline
 $\eta_{\mathrm H}$ & $<[40\!+\!\mh]^\circ$ \\
 \mhr & $[\mh\!-\!15,\mh\!+\!5]$ \\
\hline
\end{tabular}
\end{center}
\end{table}

The complete set of cuts is listed in Table~\ref{Tab:tncscuts}.
As in Ref.~\cite{HCH172}, events are accepted if they pass either 
analysis.
Typical efficiencies and background expectations are given
in Table~\ref{effbgtable}.
The main contributions to the systematic error (3\%) on the 
efficiency are: Monte Carlo
statistics; the luminosity measurement accuracy (${<} 1\%$);  
the uncertainty (${<} 2\%$) on the knowledge of the 
inefficiency introduced by beam related energy deposits at polar 
angles below 12$^\circ$, studied using events triggered at random 
beam crossings.
\par
Nine events were selected in the data for Higgs boson masses from 
40 to 75~GeV/$c^2$, in agreement with the background expectation 
of~8.7. 

For the determination of the result, the \ww\ background, 
representing 93\% of the total contamination, is reduced by a 
systematic error of 13\% and subtracted. Here the systematic error 
represents the statistical precision of a test of the \ww\ Monte Carlo 
for masses reconstructed in the region of the expected sensitivity. The test 
is performed using a data sample dominated by the \ww\  process,  
selected by relaxing the cuts on the hadronic acollinearity and on the energy 
of the leading lepton and requiring the hadronic mass to be less than 75~\gevcc{}.
In the data, 63~events were observed, in agreement with the expectation of 73.5~events.

\begin{boldmath}
\subsection{The \cscs{} final state}
\end{boldmath}

For this channel, the hadronic decays of the two charged Higgs bosons
lead to a final state with four well separated jets.
With respect to Ref.~\cite{HCH172}, the preselection applied to identify 
four-jet final states and the choice of jet pairing 
are unchanged, but the variables discriminating 
signal and Standard Model processes are exploited in a different 
way to face the larger \ww\  background. These variables are 

\begin{itemize}
\item the production polar angle $\theta_{\mathrm{prod}}$ between the Higgs 
      boson momentum direction and the beam axis;
\item the decay angles $\theta_{\mathrm{dec},i}$~($i$=1,2) in the rest frame of 
      the two reconstructed $\mathrm{H}^\pm$ candidates; 

\item the chi squared $\chi^2_{5\mathrm{C}}$ of the 5C-fit.
\end{itemize}

In addition, a $\mathrm{c}$-jet tagging variable $c_{\mathrm tag}$ is introduced 
to take advantage of the presence of $\mathrm{c}$ quarks in the signal decay products. 
This variable is the output of a neural network trained to discriminate 
$\mathrm{c}$-jets from light quark jets. 
The lifetime and specific decay modes of D mesons 
as well as jet-shape properties are exploited. A detailed description 
can be found in Ref.~\cite{VCS}.

\begin{figure}[h]
\begin{center}
\epsfig{figure=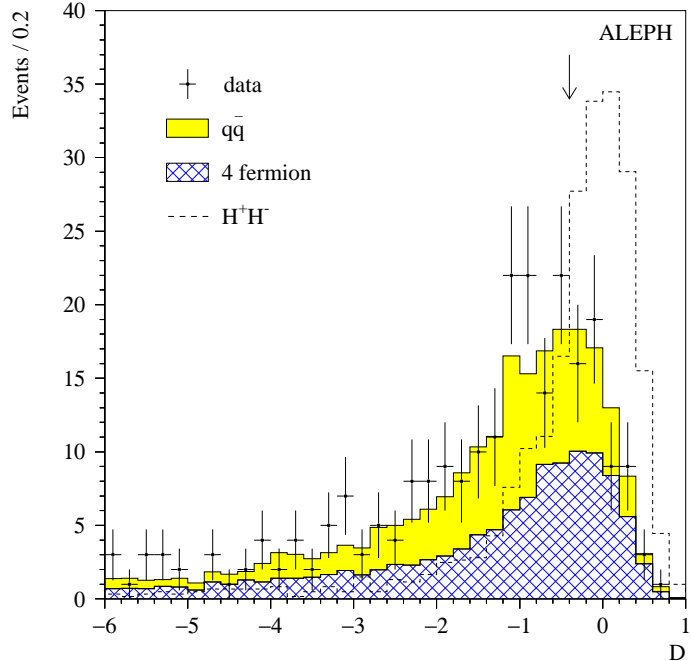,width=0.55\textwidth}
\caption{\label{Fig:hphm_fig2}
{\small The distribution of the discriminant variable $D$ used in the \cscs{} 
selection. Shown are data (points with error bars), Monte Carlo for the expected  
background sources~(cumulative histograms, normalized to the recorded 
luminosity) and the 
Monte Carlo expectation for a signal with $\mh=55$~\gevcc\ (dashed histogram,  
arbitrary normalization). 
The arrow indicates the position of the cut applied. 
Some cuts have been relaxed to preserve sufficient 
statistics.
}}
\end{center}
\end{figure}

The four variables are combined linearly into one discriminant observable:
$$D = -\cos^2\theta_{\mathrm{prod}} + 0.4\ c_{\mathrm{tag}} 
      - 0.2\ \mathrm{Min} (\cos\theta_{\mathrm{dec},i})^2 - 0.6\ \chi^2_{\mathrm{5C}}.$$
The distribution of $D$ is shown in Fig.~\ref{Fig:hphm_fig2}. 
Events are accepted if $D\geq -0.4$. 

The selection cuts are summarized in Table~\ref{Tab:cssccuts}. 
After these cuts, the total background expected for reconstructed
charged Higgs boson masses
lower than 75~\gevcc{} amounts to 42.8 events. 
Efficiencies are of the order of $35\%$ within a
dijet mass window of $\pm 3$~\gevcc{} around the Higgs boson mass hypothesis,
as shown in Table~\ref{effbgtable}. 

\begin{figure}[h]
\begin{center}
\epsfig{figure=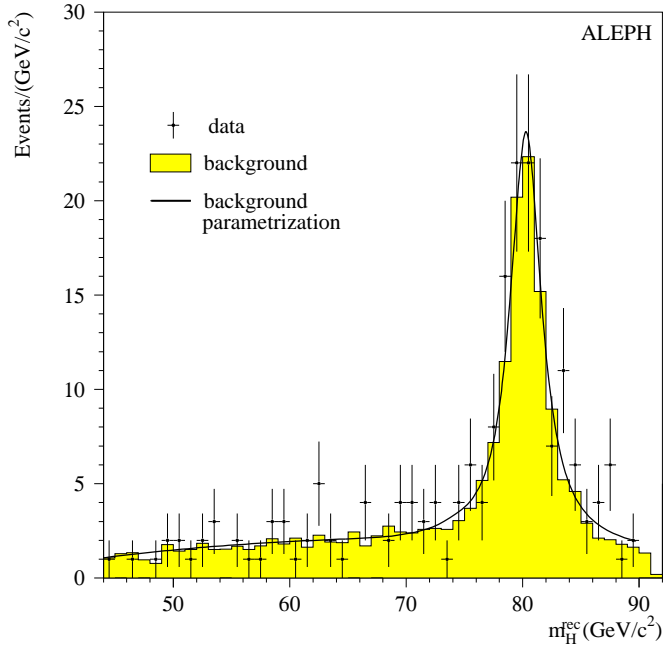,width=0.6\textwidth}
\caption{\label{m5c}{\small Distribution of the dijet mass $\mathrm{ m^{rec}_{H}}$  
obtained after applying all cuts of the \cscs{} selection; shown are 130--183~GeV data 
(dots with error bars), background Monte Carlo (shaded histogram) and the 
background parametrization used in deriving the limits (solid line).}}
\end{center}
\end{figure}

As in Ref.~\cite{HCH172}, given the level of irreducible background, 
the sensitivity
of the analysis is considerably increased by subtracting the expected
background from Standard Model processes. For this purpose
the dijet mass distribution as obtained from the background Monte Carlo
is parametrized by the sum of a polynomial and a Breit-Wigner distributions. 
The comparison with the data (Fig.~\ref{m5c}) shows that
the parametrization is consistent with the
observation. In the
following, the subtracted background is conservatively reduced
by $11\%$, corresponding to the statistical uncertainty of this comparison.

The systematic error on the number of signal events expected is 
estimated to be $2\%$, dominated
by the Monte Carlo statistical uncertainty, with
small additional contributions from the luminosity measurement and
possible inaccuracies in the simulation of the 
energy flow reconstruction.
\begin{table}[h!]
\begin{center}
\caption{\label{Tab:cssccuts}
{\small Summary of the cuts applied in the \cscs{} analysis.
Units for masses and momenta are \gevcc\ and 
\gevc, respectively. The preselection variables have been defined in Ref.~[2].  
}}
\vspace{.3cm}
\begin{tabular}{|c|c|} 
\hline
          \multicolumn{2}{|c|}{Four-jet preselection}  \\
\hline
 Good Tracks & $>7$  \\
 Charged Energy & $>10$\%\sqs \\
 $| P_{\mathrm{z-axis}} |$ & $< 1.5 [\mvis\!-\!90]$ \\
 $E^{\mathrm{em}}_{\mathrm{jet}}$ & $<90$\%$E_{\mathrm{jet}}$ \\
 $Y_{34}$ & $>0.003$  \\
 Thrust & $<0.9$  \\
\hline\hline
          \multicolumn{2}{|c|}{Equal Mass and Spin-0 constraints}  \\
 \hline
 $\theta_{\mathrm{prod}},\theta_{\mathrm{dec},i},\chi^2_{\mathrm{5C}},c_{\mathrm{tag}}$ & 
            $D\geq -0.4$ \\ 
\hline
\end{tabular}
\end{center}
\end{table}

\section{\label{combine}Results}

The number of candidate events observed  
in the data collected at centre-of-mass energies from 181 to 184~GeV
are given in Table~\ref{effbgtable} for different charged Higgs boson masses 
and for each of the three analyses presented in Section~3. 
A total of 60 events is retained for $\mh<75$~\gevcc, 
consistent with the 57.9 events expected from Standard Model processes.
Since, in addition, the mass distribution in the \cscs{} channel does 
not show any significant accumulation outside the W region~(Fig.~\ref{m5c}), 
the results of the three selections described in this note 
are combined with those obtained using 130--172~GeV data to set an improved 
95\%~C.L. lower limit on the charged Higgs boson mass, 
following the procedure described in~\cite{jadib} for the 
combination of the confidence levels and the prescription of~\cite{SHANSHIN} 
for the background subtraction.

\begin{figure}[h]
\begin{center}
\epsfig{figure=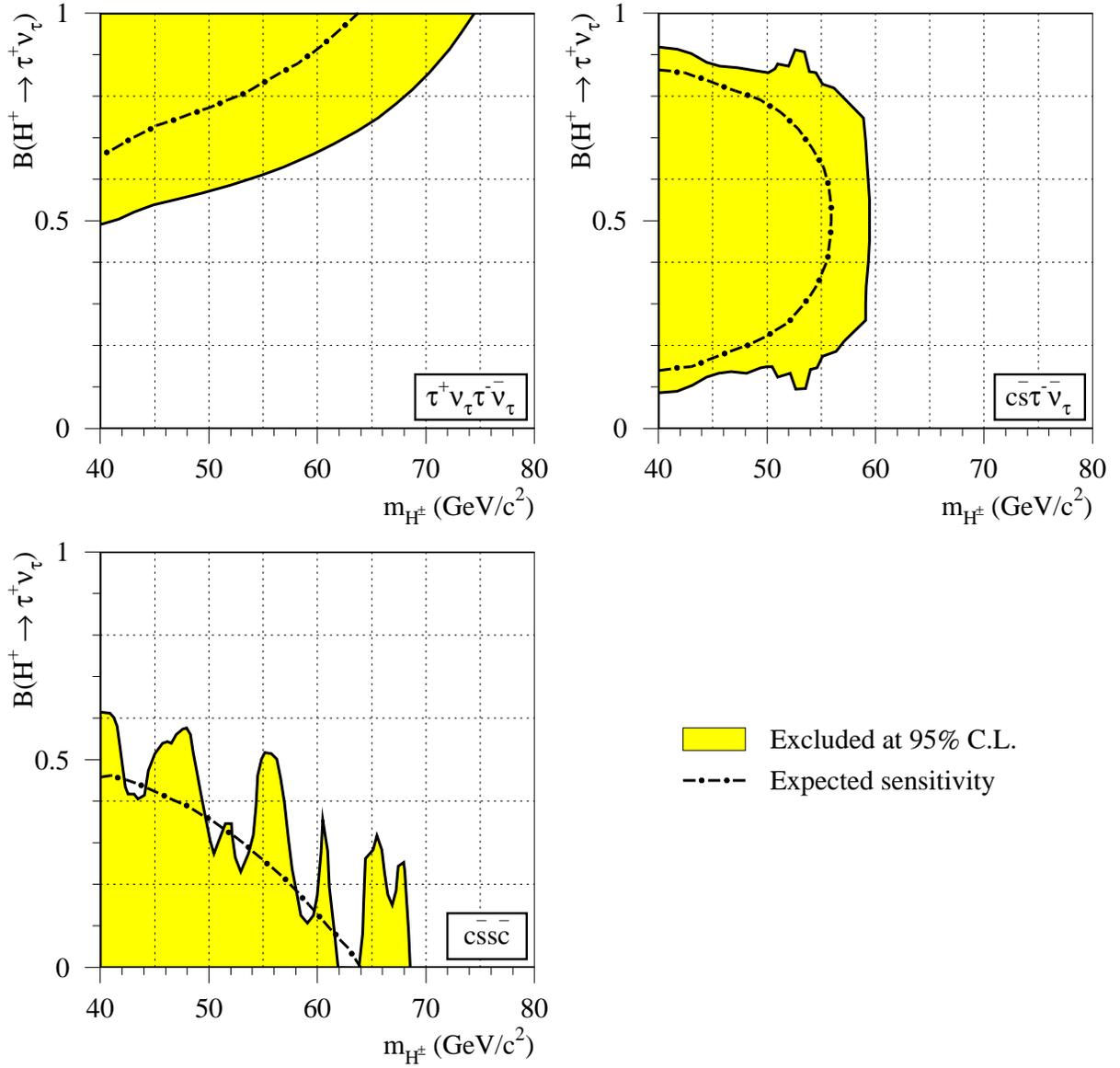,width=1.1\textwidth}
\caption{\label{separate} 
{\small Limit at $95\%$ C.L. on the mass of charged
Higgs bosons as a function
of \btn\  from each of the three individual 
analyses. Shown are the expected (dash-dotted curves) 
and observed (shaded areas) 95\% C.L. exclusion domains. 
}}
\end{center}
\end{figure}

The separate results of the three analyses are 
displayed in Fig.~\ref{separate}, where the 
contours corresponding to expected (dash-dotted curves) and observed
(solid curves) confidence levels of 5\% (equivalent
to a 95\% C.L. exclusion) are drawn.
For  $\btn=0$, 0.5 and 1, values maximizing in turn the weight of the three 
channels \cscs{}, \tncs{} and \tntn{},   
95\% C.L. lower limits on $\mh$ are set to 62, 59.5 and 
74.5~$\gevcc$.

\begin{figure}[h]
\begin{center}
\epsfig{figure=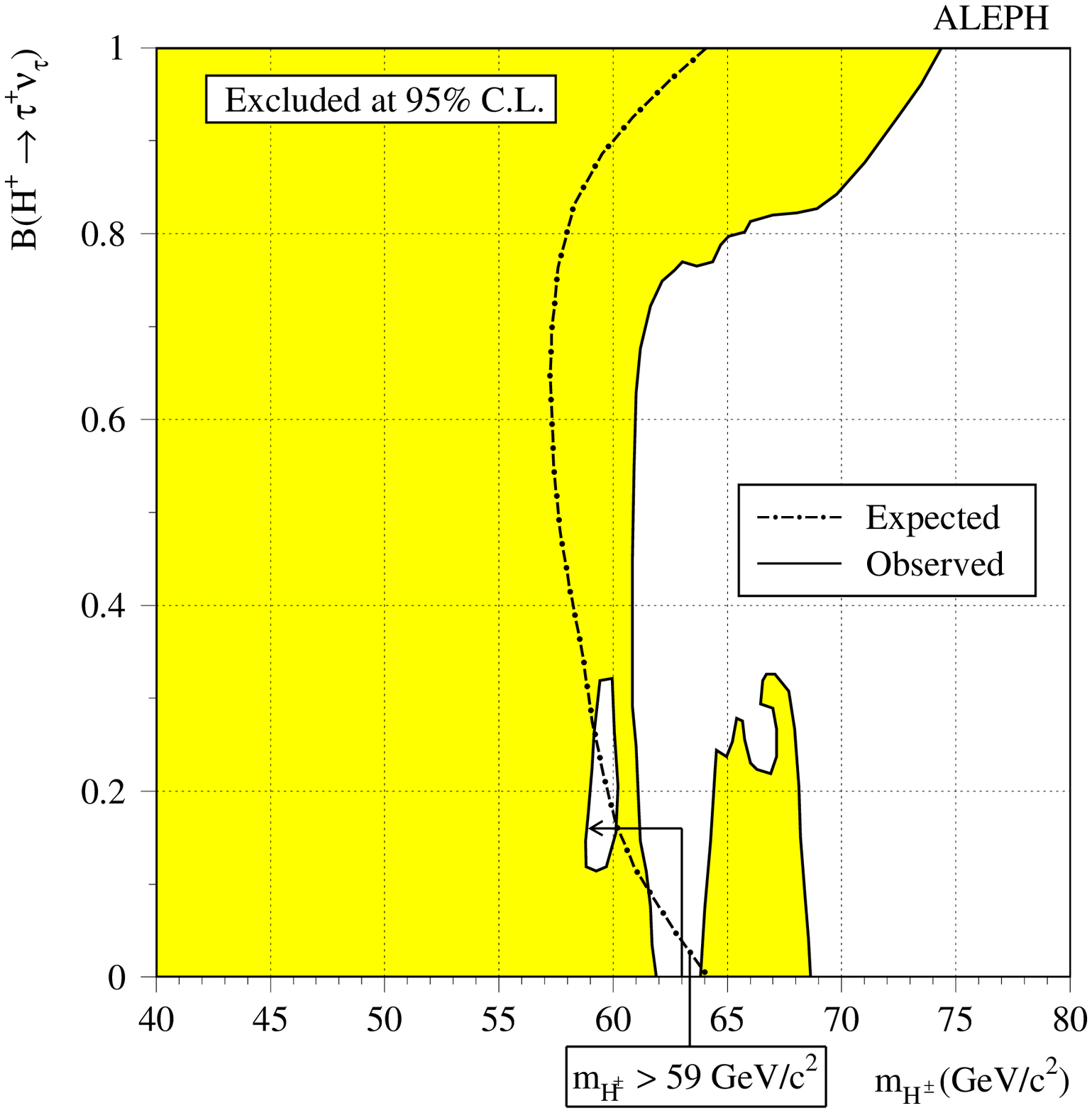,width=0.8\textwidth}
\caption{\label{combi} 
{\small Limit at $95\%$ C.L. on the mass of charged
Higgs bosons as a function
of \btn\  from the combination 
of the three individual analyses. 
 Shown are the expected (dash-dotted curve) 
and observed (shaded area) 95\% C.L. exclusion domains. 
The hole between 59 and 60~\gevcc\ and with~\btn{}$\sim0.2$ is actually 
excluded at 90\% C.L. 
}}
\end{center}
\end{figure}

The result of the combination of the three 
analyses is displayed in Fig.~\ref{combi}. 
Charged Higgs bosons with masses less than 59~\gevcc{} are excluded
at 95\% confidence level independently of \btn, 
in agreement with the 
expected exclusion sensitivity of 57 \gevcc.

\section{Conclusions}
The search for pair-produced charged Higgs bosons in the three final states
\tntn, \tncs{} and \cscs{} has been updated using 56.9~\invpb{} of data
collected at $\sqrt{s}=181-184$~GeV. No evidence of Higgs boson production
was found and new mass limits were set as a function of \btn.
When combined with data recorded at centre-of-mass energies from 130 to
172~GeV,
charged Higgs bosons with masses below 59~\gevcc{} are excluded at
$95\%$~C.L. independently of \btn.

\section{Acknowledgements}
It is a pleasure to congratulate our colleagues from the accelerator divisions
for the successful operation of $\mathtt{LEP\ 2}$. 
We are indebted to the engineers and technicians in all our institutions
for their contribution to the excellent performance of {\tt ALEPH}.
Those of us from nonmember states wish to thank {\tt CERN} for its hospitality
and support.

\end{document}

%% file: authb.tex
\pagestyle{empty}
\newpage
\small
%
%
\newlength{\saveparskip}
\newlength{\savetextheight}
\newlength{\savetopmargin}
\newlength{\savetextwidth}
\newlength{\saveoddsidemargin}
\newlength{\savetopsep}
\setlength{\saveparskip}{\parskip}
\setlength{\savetextheight}{\textheight}
\setlength{\savetopmargin}{\topmargin}
\setlength{\savetextwidth}{\textwidth}
\setlength{\saveoddsidemargin}{\oddsidemargin}
\setlength{\savetopsep}{\topsep}
%
%
\setlength{\parskip}{0.0cm}
\setlength{\textheight}{25.0cm}
\setlength{\topmargin}{-1.5cm}
\setlength{\textwidth}{16 cm}
\setlength{\oddsidemargin}{-0.0cm}
\setlength{\topsep}{1mm}
\pretolerance=10000
\textheight=25.0cm
\textwidth=16.5cm
\centerline{\large\bf The ALEPH Collaboration}
\footnotesize
\vspace{0.5cm}
{\raggedbottom
\begin{sloppypar}
\samepage\noindent
R.~Barate,
D.~Decamp,
P.~Ghez,
C.~Goy,
S.~Jezequel,
J.-P.~Lees,
F.~Martin,
E.~Merle,
\mbox{M.-N.~Minard},
\mbox{J.-Y.~Nief},
B.~Pietrzyk
\nopagebreak
\begin{center}
\parbox{15.5cm}{\sl\samepage
Laboratoire de Physique des Particules (LAPP), IN$^{2}$P$^{3}$-CNRS,
F-74019 Annecy-le-Vieux Cedex, France}
\end{center}\end{sloppypar}
\vspace{2mm}
\begin{sloppypar}
\noindent
R.~Alemany,
M.P.~Casado,
M.~Chmeissani,
J.M.~Crespo,
M.~Delfino,
E.~Fernandez,
M.~Fernandez-Bosman,
Ll.~Garrido,$^{15}$
E.~Graug\`{e}s,
A.~Juste,
M.~Martinez,
G.~Merino,
R.~Miquel,
Ll.M.~Mir,
P.~Morawitz,
A.~Pacheco,
I.C.~Park,
A.~Pascual,
I.~Riu,
F.~Sanchez
\nopagebreak
\begin{center}
\parbox{15.5cm}{\sl\samepage
Institut de F\'{i}sica d'Altes Energies, Universitat Aut\`{o}noma
de Barcelona, 08193 Bellaterra (Barcelona), E-Spain$^{7}$}
\end{center}\end{sloppypar}
\vspace{2mm}
\begin{sloppypar}
\noindent
A.~Colaleo,
D.~Creanza,
M.~de~Palma,
G.~Gelao,
G.~Iaselli,
G.~Maggi,
M.~Maggi,
S.~Nuzzo,
A.~Ranieri,
G.~Raso,
F.~Ruggieri,
G.~Selvaggi,
L.~Silvestris,
P.~Tempesta,
A.~Tricomi,$^{3}$
G.~Zito
\nopagebreak
\begin{center}
\parbox{15.5cm}{\sl\samepage
Dipartimento di Fisica, INFN Sezione di Bari, I-70126 Bari, Italy}
\end{center}\end{sloppypar}
\vspace{2mm}
\begin{sloppypar}
\noindent
X.~Huang,
J.~Lin,
Q. Ouyang,
T.~Wang,
Y.~Xie,
R.~Xu,
S.~Xue,
J.~Zhang,
L.~Zhang,
W.~Zhao
\nopagebreak
\begin{center}
\parbox{15.5cm}{\sl\samepage
Institute of High-Energy Physics, Academia Sinica, Beijing, The People's
Republic of China$^{8}$}
\end{center}\end{sloppypar}
\vspace{2mm}
\begin{sloppypar}
\noindent
D.~Abbaneo,
U.~Becker,$^{22}$
G.~Boix,$^{2}$
M.~Cattaneo,
V.~Ciulli,
G.~Dissertori,
H.~Drevermann,
R.W.~Forty,
M.~Frank,
F.~Gianotti,
A.W.~Halley,
J.B.~Hansen,
J.~Harvey,
P.~Janot,
B.~Jost,
I.~Lehraus,
O.~Leroy,
C.~Loomis,
P.~Maley,
P.~Mato,
A.~Minten,
A.~Moutoussi,
F.~Ranjard,
L.~Rolandi,
D.~Rousseau,
D.~Schlatter,
M.~Schmitt,$^{12}$
O.~Schneider,$^{23}$
W.~Tejessy,
F.~Teubert,
I.R.~Tomalin,
E.~Tournefier,
M.~Vreeswijk,
A.E.~Wright
\nopagebreak
\begin{center}
\parbox{15.5cm}{\sl\samepage
European Laboratory for Particle Physics (CERN), CH-1211 Geneva 23,
Switzerland}
\end{center}\end{sloppypar}
\vspace{2mm}
\begin{sloppypar}
\noindent
Z.~Ajaltouni,
F.~Badaud
G.~Chazelle,
O.~Deschamps,
S.~Dessagne,
A.~Falvard,
C.~Ferdi,
P.~Gay,
C.~Guicheney,
P.~Henrard,
J.~Jousset,
B.~Michel,
S.~Monteil,
\mbox{J-C.~Montret},
D.~Pallin,
P.~Perret,
F.~Podlyski
\nopagebreak
\begin{center}
\parbox{15.5cm}{\sl\samepage
\vskip -.3cm
Laboratoire de Physique Corpusculaire, Universit\'e Blaise Pascal,
IN$^{2}$P$^{3}$-CNRS, Clermont-Ferrand, F-63177 Aubi\`{e}re, France}
\end{center}\end{sloppypar}
\vspace{2mm}
\begin{sloppypar}
\noindent
J.D.~Hansen,
J.R.~Hansen,
P.H.~Hansen,
0B.S.~Nilsson,
B.~Rensch,
A.~W\"a\"an\"anen
\nopagebreak
\begin{center}
\parbox{15.5cm}{\sl\samepage
Niels Bohr Institute, 2100 Copenhagen, DK-Denmark$^{9}$}
\end{center}\end{sloppypar}
\vspace{2mm}
\begin{sloppypar}
\noindent
G.~Daskalakis,
A.~Kyriakis,
C.~Markou,
E.~Simopoulou,
A.~Vayaki
\nopagebreak
\begin{center}
\parbox{15.5cm}{\sl\samepage
Nuclear Research Center Demokritos (NRCD), GR-15310 Attiki, Greece}
\end{center}\end{sloppypar}
\vspace{2mm}
\begin{sloppypar}
\noindent
A.~Blondel,
\mbox{J.-C.~Brient},
F.~Machefert,
A.~Roug\'{e},
M.~Swynghedauw,
R.~Tanaka,
A.~Valassi,$^{6}$
H.~Videau
\nopagebreak
\begin{center}
\parbox{15.5cm}{\sl\samepage
Laboratoire de Physique Nucl\'eaire et des Hautes Energies, Ecole
Polytechnique, IN$^{2}$P$^{3}$-CNRS, \mbox{F-91128} Palaiseau Cedex, France}
\end{center}\end{sloppypar}
\vspace{2mm}
\begin{sloppypar}
\noindent
E.~Focardi,
G.~Parrini,
K.~Zachariadou
\nopagebreak
\begin{center}
\parbox{15.5cm}{\sl\samepage
Dipartimento di Fisica, Universit\`a di Firenze, INFN Sezione di Firenze,
I-50125 Firenze, Italy}
\end{center}\end{sloppypar}
\vspace{2mm}
\begin{sloppypar}
\noindent
R.~Cavanaugh,
M.~Corden,
C.~Georgiopoulos
\nopagebreak
\begin{center}
\parbox{15.5cm}{\sl\samepage
Supercomputer Computations Research Institute,
Florida State University,
Tallahassee, FL 32306-4052, USA $^{13,14}$}
\end{center}\end{sloppypar}
\vspace{2mm}
\begin{sloppypar}
\noindent
A.~Antonelli,
G.~Bencivenni,
G.~Bologna,$^{4}$
F.~Bossi,
P.~Campana,
G.~Capon,
F.~Cerutti,
V.~Chiarella,
P.~Laurelli,
G.~Mannocchi,$^{5}$
F.~Murtas,
G.P.~Murtas,
L.~Passalacqua,
M.~Pepe-Altarelli$^{1}$
\nopagebreak
\begin{center}
\parbox{15.5cm}{\sl\samepage
Laboratori Nazionali dell'INFN (LNF-INFN), I-00044 Frascati, Italy}
\end{center}\end{sloppypar}
\vspace{2mm}
\pagebreak
\begin{sloppypar}
\noindent
M.~Chalmers,
L.~Curtis,
J.G.~Lynch,
P.~Negus,
V.~O'Shea,
B.~Raeven,
C.~Raine,
D.~Smith,
P.~Teixeira-Dias,
A.S.~Thompson,
J.J.~Ward
\nopagebreak
\begin{center}
\parbox{15.5cm}{\sl\samepage
Department of Physics and Astronomy, University of Glasgow, Glasgow G12
8QQ,United Kingdom$^{10}$}
\end{center}\end{sloppypar}
\vspace{2mm}
\nopagebreak
\begin{sloppypar}
\noindent
O.~Buchm\"uller,
S.~Dhamotharan,
C.~Geweniger,
P.~Hanke,
G.~Hansper,
V.~Hepp,
E.E.~Kluge,
A.~Putzer,
J.~Sommer,
K.~Tittel,
S.~Werner,$^{22}$
M.~Wunsch
\nopagebreak
\begin{center}
\parbox{15.5cm}{\sl\samepage
Institut f\"ur Hochenergiephysik, Universit\"at Heidelberg, D-69120
Heidelberg, Germany$^{16}$}
\end{center}\end{sloppypar}
\vspace{2mm}
\begin{sloppypar}
\noindent
R.~Beuselinck,
D.M.~Binnie,
W.~Cameron,
P.J.~Dornan,$^{1}$
M.~Girone,
S.~Goodsir,
N.~Marinelli,
E.B.~Martin,
J.~Nash,
J.~Nowell,
J.K.~Sedgbeer,
P.~Spagnolo,
E.~Thomson,
M.D.~Williams
\nopagebreak
\begin{center}
\parbox{15.5cm}{\sl\samepage
Department of Physics, Imperial College, London SW7 2BZ,
United Kingdom$^{10}$}
\end{center}\end{sloppypar}
\vspace{2mm}
\begin{sloppypar}
\noindent
V.M.~Ghete,
P.~Girtler,
E.~Kneringer,
D.~Kuhn,
G.~Rudolph
\nopagebreak
\begin{center}
\parbox{15.5cm}{\sl\samepage
Institut f\"ur Experimentalphysik, Universit\"at Innsbruck, A-6020
Innsbruck, Austria$^{18}$}
\end{center}\end{sloppypar}
\vspace{2mm}
\begin{sloppypar}
\noindent
A.P.~Betteridge,
C.K.~Bowdery,
P.G.~Buck,
P.~Colrain,
G.~Crawford,
G.~Ellis,
A.J.~Finch,
F.~Foster,
G.~Hughes,
R.W.L.~Jones,
N.A.~Robertson,
M.I.~Williams
\nopagebreak
\begin{center}
\parbox{15.5cm}{\sl\samepage
Department of Physics, University of Lancaster, Lancaster LA1 4YB,
United Kingdom$^{10}$}
\end{center}\end{sloppypar}
\vspace{2mm}
\begin{sloppypar}
\noindent
P.~van~Gemmeren,
I.~Giehl,
F.~H\"olldorfer,
C.~Hoffmann,
K.~Jakobs,
K.~Kleinknecht,
M.~Kr\"ocker,
H.-A.~N\"urnberger,
G.~Quast,
B.~Renk,
E.~Rohne,
H.-G.~Sander,
S.~Schmeling,
H.~Wachsmuth
C.~Zeitnitz,
T.~Ziegler
\nopagebreak
\begin{center}
\parbox{15.5cm}{\sl\samepage
Institut f\"ur Physik, Universit\"at Mainz, D-55099 Mainz, Germany$^{16}$}
\end{center}\end{sloppypar}
\vspace{2mm}
\begin{sloppypar}
\noindent
J.J.~Aubert,
C.~Benchouk,
A.~Bonissent,
J.~Carr,$^{1}$
P.~Coyle,
A.~Ealet,
D.~Fouchez,
F.~Motsch,
P.~Payre,
M.~Talby,
M.~Thulasidas,
A.~Tilquin
\nopagebreak
\begin{center}
\parbox{15.5cm}{\sl\samepage
Centre de Physique des Particules, Facult\'e des Sciences de Luminy,
IN$^{2}$P$^{3}$-CNRS, F-13288 Marseille, France}
\end{center}\end{sloppypar}
\vspace{2mm}
\begin{sloppypar}
\noindent
M.~Aleppo,
M.~Antonelli,
F.~Ragusa
\nopagebreak
\begin{center}
\parbox{15.5cm}{\sl\samepage
Dipartimento di Fisica, Universit\`a di Milano e INFN Sezione di
Milano, I-20133 Milano, Italy.}
\end{center}\end{sloppypar}
\vspace{2mm}
\begin{sloppypar}
\noindent
R.~Berlich,
V.~B\"uscher,
H.~Dietl,
G.~Ganis,
K.~H\"uttmann,
G.~L\"utjens,
C.~Mannert,
W.~M\"anner,
\mbox{H.-G.~Moser},
S.~Schael,
R.~Settles,
H.~Seywerd,
H.~Stenzel,
W.~Wiedenmann,
G.~Wolf
\nopagebreak
\begin{center}
\parbox{15.5cm}{\sl\samepage
Max-Planck-Institut f\"ur Physik, Werner-Heisenberg-Institut,
D-80805 M\"unchen, Germany\footnotemark[16]}
\end{center}\end{sloppypar}
\vspace{2mm}
\begin{sloppypar}
\noindent
P.~Azzurri,
J.~Boucrot,
O.~Callot,
S.~Chen,
M.~Davier,
L.~Duflot,
\mbox{J.-F.~Grivaz},
Ph.~Heusse,
A.~Jacholkowska,
M.~Kado,
J.~Lefran\c{c}ois,
L.~Serin,
\mbox{J.-J.~Veillet},
I.~Videau,$^{1}$
J.-B.~de~Vivie~de~R\'egie,
D.~Zerwas
\nopagebreak
\begin{center}
\parbox{15.5cm}{\sl\samepage
Laboratoire de l'Acc\'el\'erateur Lin\'eaire, Universit\'e de Paris-Sud,
IN$^{2}$P$^{3}$-CNRS, F-91898 Orsay Cedex, France}
\end{center}\end{sloppypar}
\vspace{2mm}
\begin{sloppypar}
\noindent
G.~Bagliesi,
S.~Bettarini,
T.~Boccali,
C.~Bozzi,
G.~Calderini,
R.~Dell'Orso,
I.~Ferrante,
A.~Giassi,
A.~Gregorio,
F.~Ligabue,
A.~Lusiani,
P.S.~Marrocchesi,
A.~Messineo,
F.~Palla,
G.~Rizzo,
G.~Sanguinetti,
A.~Sciab\`a,
G.~Sguazzoni,
R.~Tenchini,
C.~Vannini,
A.~Venturi,
P.G.~Verdini
\samepage
\begin{center}
\parbox{15.5cm}{\sl\samepage
Dipartimento di Fisica dell'Universit\`a, INFN Sezione di Pisa,
e Scuola Normale Superiore, I-56010 Pisa, Italy}
\end{center}\end{sloppypar}
\vspace{2mm}
\begin{sloppypar}
\noindent
G.A.~Blair,
J.~Coles,
G.~Cowan,
M.G.~Green,
D.E.~Hutchcroft,
L.T.~Jones,
T.~Medcalf,
J.A.~Strong,
J.H.~von~Wimmersperg-Toeller
\nopagebreak
\begin{center}
\parbox{15.5cm}{\sl\samepage
Department of Physics, Royal Holloway \& Bedford New College,
University of London, Surrey TW20 OEX, United Kingdom$^{10}$}
\end{center}\end{sloppypar}
\vspace{2mm}
\begin{sloppypar}
\noindent
D.R.~Botterill,
R.W.~Clifft,
T.R.~Edgecock,
P.R.~Norton,
J.C.~Thompson
\nopagebreak
\begin{center}
\parbox{15.5cm}{\sl\samepage
Particle Physics Dept., Rutherford Appleton Laboratory,
Chilton, Didcot, Oxon OX11 OQX, United Kingdom$^{10}$}
\end{center}\end{sloppypar}
\vspace{2mm}
\begin{sloppypar}
\noindent
\mbox{B.~Bloch-Devaux},
P.~Colas,
B.~Fabbro,
G.~Fa\"if,
E.~Lan\c{c}on,
\mbox{M.-C.~Lemaire},
E.~Locci,
P.~Perez,
H.~Przysiezniak,
J.~Rander,
\mbox{J.-F.~Renardy},
A.~Rosowsky,
A.~Trabelsi,$^{20}$
B.~Tuchming,
B.~Vallage
\nopagebreak
\begin{center}
\parbox{15.5cm}{\sl\samepage
CEA, DAPNIA/Service de Physique des Particules,
CE-Saclay, F-91191 Gif-sur-Yvette Cedex, France$^{17}$}
\end{center}\end{sloppypar}
\pagebreak
\begin{sloppypar}
\noindent
S.N.~Black,
J.H.~Dann,
H.Y.~Kim,
N.~Konstantinidis,
A.M.~Litke,
M.A. McNeil,
G.~Taylor
\nopagebreak
\begin{center}
\parbox{15.5cm}{\sl\samepage
Institute for Particle Physics, University of California at
Santa Cruz, Santa Cruz, CA 95064, USA$^{19}$}
\end{center}\end{sloppypar}
\vspace{2mm}
\begin{sloppypar}
\noindent
C.N.~Booth,
S.~Cartwright,
F.~Combley,
P.N.~Hodgson,
M.S.~Kelly,
M.~Lehto,
L.F.~Thompson
\nopagebreak
\begin{center}
\parbox{15.5cm}{\sl\samepage
Department of Physics, University of Sheffield, Sheffield S3 7RH,
United Kingdom$^{10}$}
\end{center}\end{sloppypar}
\vspace{2mm}
\begin{sloppypar}
\noindent
K.~Affholderbach,
A.~B\"ohrer,
S.~Brandt,
C.~Grupen,
A.~Misiejuk,
G.~Prange,
U.~Sieler
\nopagebreak
\begin{center}
\parbox{15.5cm}{\sl\samepage
Fachbereich Physik, Universit\"at Siegen, D-57068 Siegen, Germany$^{16}$}
\end{center}\end{sloppypar}
\vspace{2mm}
\begin{sloppypar}
\noindent
G.~Giannini,
B.~Gobbo
\nopagebreak
\begin{center}
\parbox{15.5cm}{\sl\samepage
Dipartimento di Fisica, Universit\`a di Trieste e INFN Sezione di Trieste,
I-34127 Trieste, Italy}
\end{center}\end{sloppypar}
\vspace{2mm}
\begin{sloppypar}
\noindent
J.~Putz,
J.~Rothberg,
S.~Wasserbaech,
R.W.~Williams
\nopagebreak
\begin{center}
\parbox{15.5cm}{\sl\samepage
Experimental Elementary Particle Physics, University of Washington, WA 98195
Seattle, U.S.A.}
\end{center}\end{sloppypar}
\vspace{2mm}
\begin{sloppypar}
\noindent
S.R.~Armstrong,
E.~Charles,
P.~Elmer,
D.P.S.~Ferguson,
Y.~Gao,
S.~Gonz\'{a}lez,
T.C.~Greening,
O.J.~Hayes,
H.~Hu,
S.~Jin,
P.A.~McNamara III,
J.M.~Nachtman,$^{21}$
J.~Nielsen,
W.~Orejudos,
Y.B.~Pan,
Y.~Saadi,
I.J.~Scott,
J.~Walsh,
Sau~Lan~Wu,
X.~Wu,
G.~Zobernig
\nopagebreak
\begin{center}
\parbox{15.5cm}{\sl\samepage
Department of Physics, University of Wisconsin, Madison, WI 53706,
USA$^{11}$}
\end{center}\end{sloppypar}
}
\footnotetext[1]{Also at CERN, 1211 Geneva 23, Switzerland.}
\footnotetext[2]{Supported by the Commission of the European Communities,
contract ERBFMBICT982894.}
\footnotetext[3]{Also at Dipartimento di Fisica, INFN Sezione di Catania,
95129 Catania, Italy.}
\footnotetext[4]{Also Istituto di Fisica Generale, Universit\`{a} di
Torino, 10125 Torino, Italy.}
\footnotetext[5]{Also Istituto di Cosmo-Geofisica del C.N.R., Torino,
Italy.}
\footnotetext[6]{Now at LAL, 91898 Orsay, France.}
\footnotetext[7]{Supported by CICYT, Spain.}
\footnotetext[8]{Supported by the National Science Foundation of China.}
\footnotetext[9]{Supported by the Danish Natural Science Research Council.}
\footnotetext[10]{Supported by the UK Particle Physics and Astronomy Research
Council.}
\footnotetext[11]{Supported by the US Department of Energy, grant
DE-FG0295-ER40896.}
\footnotetext[12]{Now at Harvard University, Cambridge, MA 02138, U.S.A.}
\footnotetext[13]{Supported by the US Department of Energy, contract
DE-FG05-92ER40742.}
\footnotetext[14]{Supported by the US Department of Energy, contract
DE-FC05-85ER250000.}
\footnotetext[15]{Permanent address: Universitat de Barcelona, 08208 Barcelona,
Spain.}
\footnotetext[16]{Supported by the Bundesministerium f\"ur Bildung,
Wissenschaft, Forschung und Technologie, Germany.}
\footnotetext[17]{Supported by the Direction des Sciences de la
Mati\`ere, C.E.A.}
\footnotetext[18]{Supported by Fonds zur F\"orderung der wissenschaftlichen
Forschung, Austria.}
\footnotetext[19]{Supported by the US Department of Energy,
grant DE-FG03-92ER40689.}
\footnotetext[20]{Now at D\'epartement de Physique, Facult\'e des Sciences de Tunis, 1060 Le Belv\'ed\`ere, Tunisia.}
\footnotetext[21]{Now at University of California at Los Angeles (UCLA),
Los Angeles, CA 90024, U.S.A.}
\footnotetext[22]{Now at SAP AG, 69185 Walldorf, Germany}
\footnotetext[23]{Now at Universit\'e de Lausanne, 1015 Lausanne, Switzerland.}
%
%
\setlength{\parskip}{\saveparskip}
\setlength{\textheight}{\savetextheight}
\setlength{\topmargin}{\savetopmargin}
\setlength{\textwidth}{\savetextwidth}
\setlength{\oddsidemargin}{\saveoddsidemargin}
\setlength{\topsep}{\savetopsep}
\normalsize
\newpage
\pagestyle{plain}
\setcounter{page}{1}

%% file: paper_183.bbl
\clearpage

\begin{thebibliography}{99}

\bibitem{HUNTERS} J.~F.~Gunion, H.~E.~Haber, G.~Kane and S.~Dawson,
          {\it ``The Higgs Hunter's Guide''},
           Frontiers~in~Physics, Lecture~Note~Series,
           Addison~Wesley, 1990.

\bibitem{HCH172} ALEPH Collaboration, {\it ``Search for charged Higgs 
bosons in $e^+e^-$ collisions at centre-of-mass energies 
from 130 to 172 GeV''}, Phys. Lett. {\bf B 418}~(1998)~419.

\bibitem{DELPHI172} DELPHI Collaboration, 
{\it ``Search for charged Higgs bosons in $e^+e^-$ collisions at 
$\sqrt{s}=$~172~GeV''},
Phys. Lett. {\bf B 420}~(1998)~140.

\bibitem{L3183} L3 Collaboration, 
{\it ``Search for Charged Higgs Bosons in $e^+e^-$  Collisions at Centre-of-Mass 
Energies between 130 and 183~GeV''}, 
CERN-EP/98-149~(1998),
submitted~to~Phys.~Lett.~B.

\bibitem{OPAL183} OPAL Collaboration, 
{\it ``Search for Higgs bosons in $e^+e^-$ Collisions at 183~GeV''},
CERN-EP/98-173~(1998),
submitted~to~Eur.~Phys.~J.~C.

\bibitem{btau} ALEPH Collaboration, 
{\it ``Measurement of the $b\!\to\!\tau^- \bar{\nu}_\tau X$ branching ratio and 
an upper limit on $B^-\!\to\! \tau^-\bar{\nu}_\tau$''}, 
Phys. Lett. {\bf B 343} (1995) 444.

\bibitem{cleo} CLEO Collaboration, 
{\it ``First measurement of the rate for the inclusive radiative penguin 
decay $b\!\to\!s\gamma$''}, 
Phys.~Rev.~Lett.~{\bf 74}~(1995)~2885.

\bibitem{cdf.higgs} CDF Collaboration, {\it ``Search for Charged Higgs
Decays of the Top Quark using Hadronic Decays of the Tau Lepton''},
Phys.~Rev.~Lett.~{\bf 79}~(1997)~357.

\bibitem{bib:detectorpaper} ALEPH Collaboration, 
{\it ``ALEPH: a detector for electron-positron annihilations at LEP''}, 
Nucl. Instrum. and Methods {\bf A 294}~(1990) 121.

\bibitem{bib:performancepaper} ALEPH Collaboration, 
{\it ``Performance of the ALEPH detector at LEP''},
Nucl. Instrum. and Methods {\bf A 360} (1995) 481.


\bibitem{nbar95} J.-F.~Grivaz and F.~Le~Diberder, {\it ``Complementary
Analyses and Acceptance Optimization in new Particle Searches''},
LAL~preprint~\#~92-37~(1992).

\bibitem{WAS1} S.~Jadach, B.F.L.~Ward, and Z.~W\c{a}s,
{\it ``The Monte Carlo program KORALZ, version 4.0, for the lepton or quark pair 
production at LEP/SLC energies''}, 
Comp.~Phys.~Commun.~{\bf 79}~(1994)~503.
\bibitem{PYTHIA} T.~Sj\"ostrand,
                 {\it ``The~PYTHIA~5.7~and~JETSET~7.4~Manual''},
                 LU-TP~95/20,~CERN-TH~7112/93,
                 Comp.~Phys.~Commun.~{\bf 82}~(1994)~74.
\bibitem{KORALW} M.~Skrzypek, S.~Jadach, W.~Placzek and Z.~W\c{a}s,
{\it ``Monte Carlo program KORALW-1.02 for W pair production at 
LEP-2/NLC energies with Yennie-Frautschi-Suura exponentiation''};
Comp.~Phys.~Commun.~{\bf 94}~(1996)~216.
\bibitem{JANOT} G.~Ganis~and~P.~Janot,
                {\it ``The~HZHA~Generator''}
                in {\it ``Physics~at~LEP2''}, 
                Eds. G.~Altarelli, T.~Sj\"ostrand and F.~Zwirner,
                CERN~96-01~(1996),~Vol.~2, 309.

\bibitem{slepton183} ALEPH Collaboration, 
{\it ``Search for sleptons in $e^+e^-$ collisions at 
centre-of-mass energies up to 184 GeV''}, 
Phys. Lett. {\bf B 433}~(1998)~176.

\bibitem{VCS} ALEPH Collaboration, 
{\it ``Measurement of $| V_{cs} |$ in hadronic W decays''},
 ALEPH 98-011, CONF 98-001, available from 
{\tt http://alephwww.cern.ch/ALPUB/}.



\bibitem{jadib} P.~Janot and F.~Le~Diberder, {\it ``Optimally combined 
confidence levels''},
Nucl. Instrum. and Methods {\bf A 411}~(1998)~449.

\bibitem{SHANSHIN} S. Jin and P. McNamara, {\it ``The Signal Estimator Limit 
Setting Method''}, {\tt physics/9812030}, submitted to Nucl. Instrum. and 
Methods~A.


\end{thebibliography}
